\documentclass[10pt,epsf]{article}

\usepackage{amsmath}
\usepackage{narosa}
\usepackage{hhline}

\usepackage{graphicx}
\usepackage{dcolumn}
 \usepackage{setspace}
\usepackage{bm}
\def\phd#1#2#3{{ Physica } {\bf #1D}, (#2) #3}
\def\prl#1#2#3{{ Phys. Rev. Lett.} {\bf #1}, (#2) #3}
\def\pla#1#2#3{{ Phys. Lett. } {\bf #1A}, (#2) #3}
\def\pra#1#2#3{{ Phys. Rev. A} {\bf #1}, (#2) #3}
\def\pre#1#2#3{{ Phys. Rev. E} {\bf #1}, (#2) #3}

\def\ijbc#1#2#3{{ Int. J. Bifurcation and Chaos} {\bf #1}, (#2) #3}

\def\ep{\epsilon~}
\def\ie{i.e.~}

\def\etl{$et ~al.$~}

\def\ep{\epsilon}
\def\eqn{\end{equation}\noindent}
\def\eqnr{\end{eqnarray}\noindent}
\def\bc{\begin{center}}
\def\ec{\end{center}}

\begin{document}

\title[SNAs]{Strange nonchaotic attractors in driven delay--dynamics}

\author{Awadhesh Prasad$^\dagger$},
\author{Manish Agrawal$^\dagger$}, and
\author{Ramakrishna Ramaswamy}$^\ddagger$

\address{$^\dagger$ Department of Physics and Astrophysics,
University of Delhi, Delhi 110 007, India}

\address{$^\ddagger$  School of Physical Sciences,
Jawaharlal Nehru University, New Delhi 110 067, India}

\begin{abstract}

Strange nonchaotic attractors (SNAs) are observed in quasiperiodically driven time--delay systems. Since the largest Lyapunov exponent is nonpositive, trajectories in two such identical but distinct systems  show the property of {\it phase}--synchronization. Our results are illustrated in the model SQUID and R\"ossler oscillator systems.   

\end{abstract}

\section{Introduction}
Extensive  studies over the past twenty or so years on strange nonchaotic attractors \cite{gopy} have helped to
 establish that such behavior is not exceptional \cite{ijbc,ijbc2,book}. By now several examples are known where
 it may  be clearly shown that the dynamics is both strange and
nonchaotic \cite{gopy,keller,indu,ott}, namely that the attractor has a fractal geometry,
 and that the largest Lyapunov exponent is either zero or negative.  A number of questions relating to the
origins of such dynamics remain open. Most studies (analytical or numerical) of SNA dynamics  have been on
low--dimensional quasiperiodically driven systems. One open question relates to the need for external forcing.
All known examples of systems where SNAs occur have a skew--product structure,
and it is still not entirely clear whether this feature is necessary or merely sufficient.
 Further,  in all known examples the external drive is quasiperiodic in time:  it is also not clear
whether this form of drive is necessary for the creation of strange nonchaotic motion or merely sufficient \cite{ijbc2}.
                                                                                                                    
For chaotic attractors, there can in principle be several positive
Lyapunov exponents (the phenomenon of hyperchaos). Although the SNAs are
 geometrically similar to chaotic attractors, any further analogy between them
 is clearly not possible, and thus it is of interest to investigate the
nature of SNAs in high--dimensional dynamical systems.
                                                                                                                    
In the present work we consider model time--delay quasiperiodically driven dynamical systems and study the nature of                                                                                                                    
the dynamics. The inclusion of time--delay through a diffusive self-feedback coupling term makes the system
effectively infinite--dimensional. The SNAs that are created, however, continue to be
low--dimensional \cite{lowd}
 and are very similar to those created in the absence of time--delay seen in earlier studies \cite{zhou}. A second motivation for the present study is practical. In order to create SNAs in an experiment, the physical set--up may involve time--delays and thus it is important to know the effects of  including such coupling.  Time--delay coupling has attracted considerable interest since this can cause interesting dynamical phenomena such as amplitude  death \cite{reddy} or novel bifurcations  \cite{ap,pyra} .
                                                                                                                    
Since all Lyapunov exponents on SNAs are nonpositive, a distinctive property of such attractors is that trajectories with different initial conditions do not separate from each other. Indeed, on two identical but separate systems, trajectories started with the same phase  will completely synchronize with each other \cite{rr}. If the initial phases are different, then trajectories show phase--synchronization  \cite{phase}. Similarly, on a given SNA, trajectories starting from different initial conditions also coincide or phase--synchronize in this manner, which is one of our aim in this paper.
                                                                                                                    
In the next section of this paper we study the occurrence of SNAs in time--delay driven dynamical systems. This is followed by 
a discussion of phase synchronization in Section III. Finally, a brief summary is given in Section IV.
                                                                                                                    
\section{Delay SNAs}
We study two model quasiperiodically modulated systems.  Among the earliest demonstrations of dynamical systems with SNA  was the driven and damped pendulum equation  \cite{zhou} used to model a driven SQUID with inertia and damping. We introduce a delay feedback term, to get the equation of motion as
\begin{eqnarray}
\nonumber
{\ddot {x}}+\gamma \dot{x} +x+q_2 \sin 2\pi x &=&q_1 \sin(\omega_1 t)+\beta \sin(\omega_2 t)\\
& &+ \ep [x(t-\tau)-x(t)].
\label{eq:squid}
\end{eqnarray}
The drive is made quasiperiodic by  requiring the ratio of the frequencies to be an
irrational number. Here  we take this ratio as  inverse golden mean,
\ie $\omega_1/\omega_2= \omega=(\sqrt{5}-1)/2$.
                                                                                                                    
In a similar manner, we also consider a driven R\"ossler system \cite{rossler} with delay feedback and quasiperiodic parameter modulation,
\begin{eqnarray}
\nonumber
 {\dot {x}}&=&-y-z\\
\nonumber
 {\dot {y}}&=&x+\alpha (1+\frac{1}{2} (\cos t+\cos\omega t)) y\\
\nonumber
& & + \ep [y(t-\tau)-y(t)]\\
 {\dot {z}}&=&0.1+z (x-14).
\label{eq:ross}
\end{eqnarray}
                                                                                                                    
The above dynamical systems are integrated using standard techniques \cite{rk4}
and we compute the largest several Lyapunov exponents \cite{farmer} as a function of parameters. For a wide range of parameter values, we find that the dynamics is on nonchaotic attractors in both systems, and representative results are shown in
 Fig. 1. 

The largest Lyapunov exponent is shown in  Figs. 1 (a) and (b) for the SQUID and R\"ossler systems respectively. (See the caption for details of the parameter values).  The symbols $T$ and $C$ denote quasiperiodic torus and chaotic attractors respectively, and $AD$ indicates the amplitude--death region, where oscillations die to a  fixed point \cite{ap}. The corresponding variance of finite--time Lyapunov exponents, which has been used as an order parameter for detecting the torus to  SNA transition (see Ref. \cite{pre} for details), is shown in Figs. 1 (c) and (d) and this suggests the presence of SNAs in the region marked by arrows. Examples of such nonchaotic attractors are shown in Figs. 1 (g, h) for the two cases. The route to SNAs   appears to be via fractalization \cite{ijbc,book} (see Figs. 1 (e, f)); in this transition, a torus attractor gets wrinkled progressively as parameters are increased and
transform to  SNAs \cite{fractal} with a slow increase in the variance of finite--time  Lyapunov exponents as shown
in   Fig. 1 (c, d).
                                                                                                                    
A number of  measures \cite{pfc,ijbc2,scaling} have been suggested  in order to quantitatively confirm that the attractors are indeed SNAs. We  compute the partial Fourier sum \cite{scaling},
\begin{equation}
T(\Omega, N) = \sum_{k=1}^N x_k \exp (i2\pi k \Omega)
\end{equation}
where $\Omega$ is proportional to the irrational driving frequency
$\omega$ \cite{scaling} and $\{x_n\}$ is the time series of one of the dynamical variables. The
graph of Re $T$ vs.  Im $T$ gives a ``walk'' on the plane, with mean square displacement$
\vert T(\Omega, N)\vert^2$.  The singular--continuous nature of the SNA spectrum implies \cite{pfc} the scaling $ \vert T(\Omega, N) \vert^2 \sim N^{\mu}$ with $1 < \mu < 2$. Plots of
 $\log|T(\Omega,N)|^2$ versus $\log N$ in Figs. 1 (i,j) for the SNA dynamics corresponding
to Figs. 1 (g,h),  show   linear behavior, with slopes $\mu \approx 1.67$ (for the SQUID system) and  $\mu \approx 1.59$ (R\"ossler system), suggesting that both the attractors are indeed strange and the dynamics is nonchaotic.
                                                                                                                    
\begin{figure}[!ht]
\begin{center}
\includegraphics[width=0.45\linewidth]{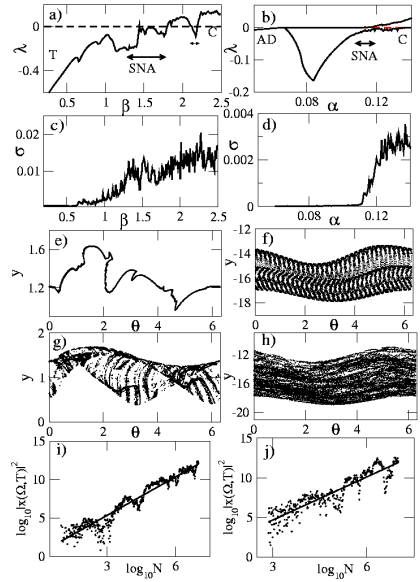}
\end{center}
\caption{ Results for the SQUID system with parameters $\tau=2$, coupling strength $\ep=0.01$,  $q_1=2.768, q_2=\gamma=2$ and
the R\"ossler system with  $\tau=0.9$. The Lyapunov exponents are plotted as a function of drive amplitude $\beta$ for the SQUID system in  (a)  and as a function of $\alpha$ in the R\"ossler system in  (b).
The variance in the Lyapunov exponents are shown in (c) \& (d) for the cases (a) \& (b) respectively. The
representative Poincare sections of periodic torus and corresponding SNAs are shown for the two systems in
  (e) $\beta=1$ $(y=\dot {x})$, \& (f) $\alpha=0.1$, and (g) $\beta=1.6$ \& (h) $\alpha=0.114$. Here
the Poincare section are taken at $x=0$, and then $\theta=mod (t, 2\pi)$ are considered.
Shown in (i) and (k) are the ``walk'' displacements, $T(\Omega,N)$ vs $N$ on a logarithmic scale for
 SNAs corresponding to (g) and (h) respectively at $\Omega=\omega/4$.  }
\label{fig:traj}
\end{figure}

\begin{figure}[!ht]
\begin{center}
\includegraphics[width=0.45\linewidth]{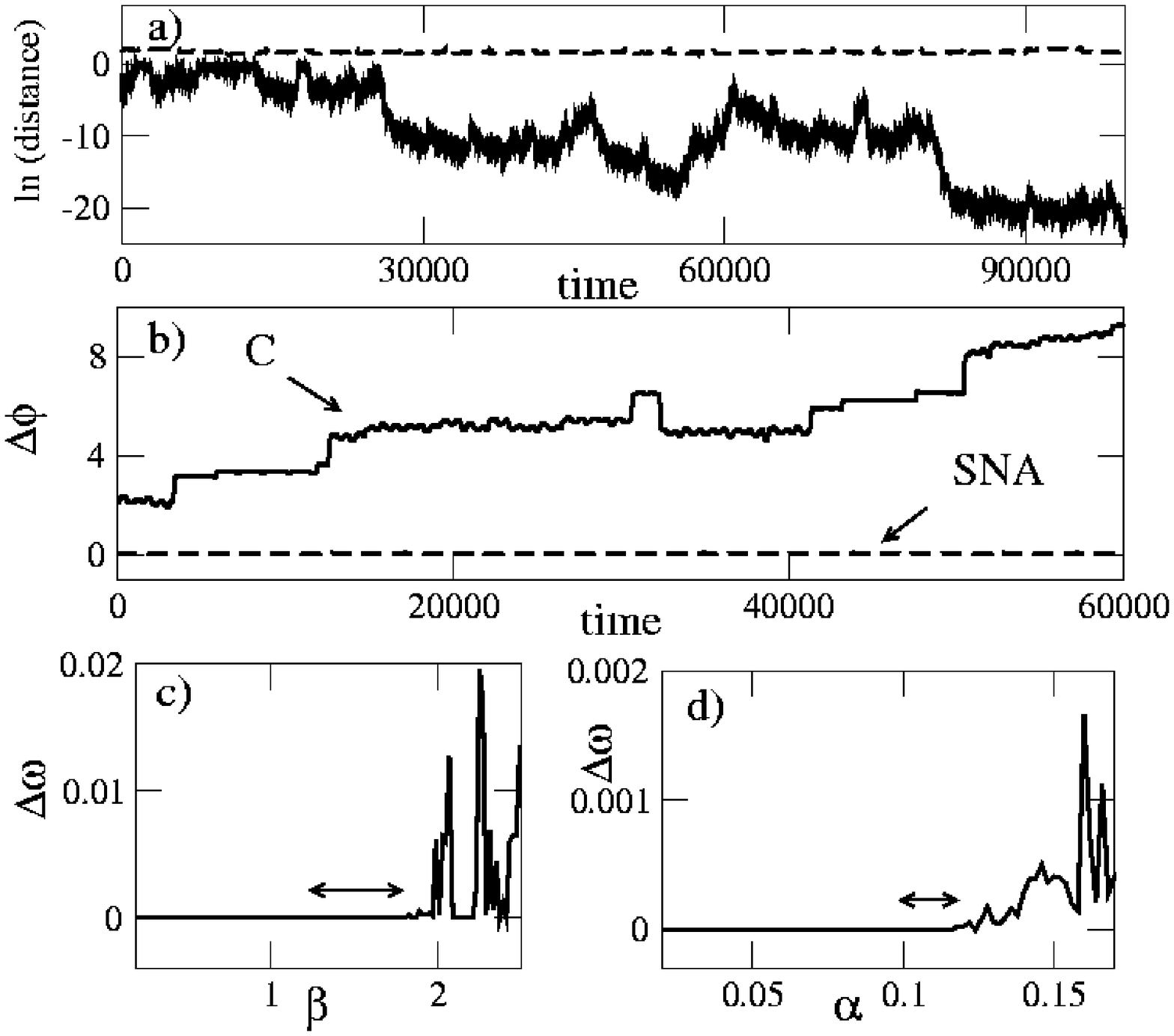}
\end{center}
\caption{ (a) Distance between  two trajectories starting with different initial conditions
 for SQUID, Eq. (\ref{eq:squid}) (solid line) and R\"ossler, Eq.
(\ref{eq:ross}) (dashed line) in SNA region at parameters $\beta=1.6$ and $\alpha=0.114$ respectively.
(b) The phase difference $\Delta \phi$ for a chaotic attractor (solid line) at $\alpha=0.2$ and SNA
 (dashed line) at $\alpha=0.114$ for R\"ossler oscillator.
(c) and (d) show the  frequency differences for SQUID and R\"ossler systems respectively. Arrows
 indicates the regimes of SNAs. The curves in (c)-(d) are averaged over $50$ random
 initial conditions sampled uniformly.}
\label{fig:phase}
\end{figure}

\section{Synchronization}
Since the largest Lyapunov exponent is negative, trajectories on SNAs  have the property that they eventually coincide and become identical (typically with a time--shift). Explicitly, for two different initial conditions denoted ${\bf x (0)}$ and ${\bf x^{\prime} (0)}$, the distance
\begin{equation}
||{\bf x (t)} - {\bf x^{\prime} (t + \tau)} || \to  0
\end{equation}
for some $\tau$. When it is possible to define a phase variable, then this is similar to a phase--shift.
                                                                                                                    
Consider now two identical systems where the dynamics is on SNAs. Trajectories starting from arbitrary initial conditions will therefore show synchronization. If the initial phases are the same, the synchronization will be complete
\cite{rr}, but more generally, this will be a phase--synchronization \cite{phase}. Since this phenomenon occurs in the absence of any coupling between the two systems, this notion of synchronization also applies to two trajectories
in the same system.

The distance between two trajectories with random initial conditions  is shown as a function of time in Fig. 2 (a) (the solid line) for  the SQUID system. This quantity rapidly goes to zero. For the R\"ossler case, earlier work has
shown that it is possible to define a ``phase'' as $\phi \sim \tan^{-1}(y/x)$ \cite{phase}. For two trajectories on
SNAs, the phase difference $\Delta \phi=\phi_1-\phi_2=\Delta \omega t$ does not grow with time (see Fig. 2 (b)) unlike what happens in the case of very morphologically similar chaotic R\"ossler attractors.  As shown in Figs. 2 (c)-(d) on SNAs  (namely in the parameter range indicated by arrows ) the motion is phase locked in both the SQUID and R\"ossler systems.

\section{Summary}
In the present work we have studied representative nonlinear time--delay 
dynamical systems with quasiperiodic forcing and observe that the dynamics 
can be on strange nonchaotic attractors. These attractors are low--dimensional 
and trajectories on these SNAs have the property of phase--synchronization. 
We have verified that these features are shared by other delay systems
such as the  Mackey--Glass equations \cite{blood},  or the time--delay Duffing oscillator
when subject to an external quasiperiodic drive.
                                                                                                                 
\bc
{\large Acknowledgment}
\ec
This work is supported by the Department of Science and Technology, Govt. of India. We have great pleasure 
in dedicating this article to Prof. M Lakshmanan on the occasion of his sixtieth birthday.


\begin{thebibliography}{22}
\bibitem{lowd} For smaller coupling strength in the feedback term, the attractor is similar to that in the absence of  feedback (which is low-dimensional in any case).
\bibitem{rk4} The flow is integrated using a  Runge-Kutta 4th order scheme  with integration step $\Delta t=\tau/N$ with $N=200$.
\end{thebibliography}
\end{document}